\documentclass[%
twocolumn,
superscriptaddress,
 amsmath,amssymb,
 aps,
 prd,
]{revtex4-2}

\usepackage{graphicx}
\usepackage{dcolumn}
\usepackage{bm}
\usepackage{hyperref}
\usepackage[makeroom]{cancel}
\usepackage{xcolor}

\def\dm{$\mathrm{DM}$}
\def\wimps{$\mathrm{WIMPs}$}
\def\NEON{$\mathrm{NEON}$}

\def\CEvNS{$\mathrm{CE\nu NS}$}
\def\Geant{$\mathrm{Geant}$}
\begin{document}


\title{First Direct Search for Light Dark Matter Using the NEON Experiment at a Nuclear Reactor}

\author{J.~J.~Choi}
\email{jjchoi1375@gmail.com}
\affiliation{Department of Physics and Astronomy, Seoul National University, Seoul 08826, Republic of Korea}
\affiliation{Center for Underground Physics, Institute for Basic Science (IBS), Daejeon 34126, Republic of Korea}
\author{C.~Ha}
\affiliation{Department of Physics, Chung-Ang University, Seoul 06973, Republic of Korea}
\author{E.~J. Jeon}
\affiliation{Center for Underground Physics, Institute for Basic Science (IBS), Daejeon 34126, Republic of Korea}
\author{J.~Y. Kim}
\affiliation{HANARO Utilization Division, Korea Atomic Energy Research Institute (KAERI), Daejeon 34057, Republic of Korea}
\author{K.~W.~Kim}
\author{S.~H.~Kim}
\affiliation{Center for Underground Physics, Institute for Basic Science (IBS), Daejeon 34126, Republic of Korea}
\author{S.~K.~Kim}
\affiliation{Department of Physics and Astronomy, Seoul National University, Seoul 08826, Republic of Korea}
\author{Y.~D.~Kim}
\affiliation{Center for Underground Physics, Institute for Basic Science (IBS), Daejeon 34126, Republic of Korea}
\affiliation{IBS School, University of Science and Technology (UST), Deajeon 34113, Republic of Korea}
\author{Y.~J.~Ko}
\email{yjko@ibs.re.kr}
\affiliation{Center for Underground Physics, Institute for Basic Science (IBS), Daejeon 34126, Republic of Korea}
\author{B.~C.~Koh}
\affiliation{Department of Physics, Chung-Ang University, Seoul 06973, Republic of Korea}
\author{S.~H.~Lee}
\affiliation{IBS School, University of Science and Technology (UST), Deajeon 34113, Republic of Korea}
\affiliation{Center for Underground Physics, Institute for Basic Science (IBS), Daejeon 34126, Republic of Korea}
\author{I.~S.~Lee}
\affiliation{Center for Underground Physics, Institute for Basic Science (IBS), Daejeon 34126, Republic of Korea}
\author{H.~Lee}
\affiliation{IBS School, University of Science and Technology (UST), Deajeon 34113, Republic of Korea}
\affiliation{Center for Underground Physics, Institute for Basic Science (IBS), Daejeon 34126, Republic of Korea}
\author{H.~S.~Lee}
\email{hyunsulee@ibs.re.kr}
\affiliation{Center for Underground Physics, Institute for Basic Science (IBS), Daejeon 34126, Republic of Korea}
\affiliation{IBS School, University of Science and Technology (UST), Deajeon 34113, Republic of Korea}
\author{J.~S.~Lee}
\author{Y.~M.~Oh}
\affiliation{Center for Underground Physics, Institute for Basic Science (IBS), Daejeon 34126, Republic of Korea}
\author{B.~J.~Park}
\affiliation{IBS School, University of Science and Technology (UST), Deajeon 34113, Republic of Korea}
\affiliation{Center for Underground Physics, Institute for Basic Science (IBS), Daejeon 34126, Republic of Korea}

\collaboration{NEON Collaboration}

\date{\today}

\begin{abstract}
We report new results from the Neutrino Elastic Scattering Observation with NaI (NEON) experiment in the search for light dark matter (LDM) using 2,636\,kg$\cdot$days of NaI(Tl) exposure.
The experiment employs an array of NaI(Tl) crystals with a total mass of 16.7\,kg, located 23.7 meters away from a 2.8\,GW thermal power nuclear reactor.
We investigated LDM produced by the \textit{invisible decay} of dark photons, a well-motivated mechanism generated by high-flux photons during reactor operation.  
The energy spectra collected during reactor-on and reactor-off periods were compared within the LDM signal region of 1--10\,keV.
No signal consistent with LDM interaction with electrons was observed, allowing us to set 90\% confidence level exclusion limits on the dark matter-electron scattering cross-section ($\sigma_e$) across dark matter masses ranging from 1\,keV/c$^2$ to 1000\,keV/c$^2$.
Our results set a 90\% confidence level upper limit of $\sigma_e = 3.17\times10^{-35}\,\mathrm{cm^2}$ for a dark matter mass of 100\,keV/c$^2$, marking the best laboratory result in this mass range.
Additionally, our search extends the coverage of LDM below 100\,keV/c$^2$ for the first time, assuming the specific \textit{invisible decay} of dark photons.

\end{abstract}

\maketitle


Astrophysical observations strongly suggest the existence of dark matter~(DM), which is non-baryonic and non-luminous~\cite{planck2020}.
Weakly Interacting Massive Particles~(\wimps) have long been considered the primary candidates for DM~\cite{PhysRevLett.39.165,Goodman:1984dc}. However, despite extensive experimental efforts, no conclusive evidence for WIMPs  has been found~\cite{Undagoitia_2015, Schumann_2019, ParticleDataGroup:2022pth}.
This has led to increased interest in alternative DM candidates. One such candidate is light dark matter~(LDM;\,$\chi$), which has masses ranging from keV/c$^2$ to GeV/c$^2$. These masses are typically lower than those of \wimps. LDM is hypothesized to exist in a dark sector that interacts with the Standard Model~(SM) sector via a mediator~\cite{Pospelov_2008, Feng_2008, PhysRevD.86.056009, Knapen_2017}.

The dark photon~(DP;\,$A'$) is one of the simplest hypothetical particles in the dark sector and can act as a mediator in the form of a massive $U(1)_D$ gauge boson~\cite{HOLDOM1986196}.
Dark photons kinetically mix with SM photons through a mixing parameter~($\varepsilon$), prompting various experimental searches for DPs~\cite{Filippi_2020}. 
Nuclear reactors are intense sources of photons with energies up to a few MeV, making them promising environments to study DPs due to their high photon flux~\cite{deNiverville_2021}. Previous phenomenological studies with NEOS and TEXONO experiments have explored DP interactions with SM particles~\cite{PhysRevLett.119.081801, Danilov_2019}. Additionally, pair produced LDMs from DPs interacting with orbit electrons in the detector material were studied in Ref.~\cite{Ge_2018}, assuming $m_{A'}=3m_{\chi}$, where $m_{A'}$ is the DP mass and $m_{\chi}$ is the LDM mass. The COHERENT experiment has investigated LDM via similar processes, focusing on interactions with target nuclei~\cite{COHERENT:2021pvd}.

This letter presents the first direct search for keV/c$^2$-scale LDM in a nuclear reactor using data from the NEON experiment. We considered scenarios where DPs decay into LDM particles, assuming $m_{A'}=3m_{\chi}$, similar to \textit{invisible decay} channels studied by beam dump experiments~\cite{Batell_2014, Marsicano_2018, andreev2023search} and the COHERENT experiment~\cite{COHERENT:2021pvd}. 
Although both fermionic and scalar DM can be considered~\cite{PhysRevD.85.063503}, we focused on fermionic DM scattering off electrons in the detector, following the methodology outlined in Ref.~\cite{Ge_2018}.

The Neutrino Elastic Scattering Observation with NaI~(NEON) experiment aims to detect coherent elastic neutrino-nucleus scattering~(\CEvNS) using reactor antineutrinos~\cite{Choi_2023}.
The NEON detector has been operational in the tendon gallery of the Hanbit Nuclear Power Plant Unit 6 since December 2020, with physics operation commencing in April 2022 following an upgrade of the detector encapsulation~\cite{choi2024upgrade,park2024}. The detector is located 23.7 meters from the reactor core.

The NEON detector consists of four 8-inch long and two 4-inch long, 3-inch diameter NaI(Tl) crystals, with a total mass of 16.7\,kg.
These six crystals are submerged in 800\,liters of a linear alkylbenzene-based liquid scintillator~(LS) contained within a acrylic box. 
The LS serves to reduce the radioactive background affecting the NaI(Tl) crystals by tagging and shielding radiations~\cite{Adhikari:2020asl}. To further minimize the external radiation background, the LS is enclosed by a shielding structure composed of 15 cm thick lead, 2.5 cm thick borated polyethylene, and 30 cm thick high-density polyethylene~\cite{Choi_2023}. 

Each NaI(Tl) crystal is coupled directly to two photomutiplier tubes~(PMTs) without any intervening windows to enhance light collection efficiency~\cite{choi2020,choi2024upgrade}. An event is triggered when coincident single photoelectrons are detected in both PMTs coupled to a single crystal within a 200\,ns time window. The signals from the PMTs are processed by 500\,MHz flash analog-to-digital converters, producing 8\,$\mu$s long waveforms that begin 2.4\,$\mu$s before the trigger.
Data acquisition is conducted through two readout channels: a high-gain anode channel for energies ranging from 0 and 60\,keV and a low-gain dynode channel for energies between 60 and 3000\,keV.
The LS signals are processed by charge-sensitive flash analog-to-digital converters.
A similar data acquisition system has been used in the COSINE-100 experiment~\cite{Adhikari_2018} without any issues for over 6\,years of operation.

To reject unwanted phosphorescence events from direct muon hits, a 300\,ms dead time is applied to muon candidate events that have energies above 3000\,keV. This results in approximately 5\% dead time for small crystals and 10\% dead time for large crystals. The rate of muon candidate events is monitored, and the dead time is evaluated for each hourly dataset. 

The distribution of the $\gamma$-ray flux from a nuclear reactor is approximated based on the FRJ-1 research reactor~\cite{Bechteler1983}:
\begin{equation}
    \frac{dN_{\gamma}}{dE_{\gamma}}=0.58\times10^{21}\left(\frac{P}{\mathrm{GW}}\right)\mathrm{exp}\left(-\frac{E_{\gamma}}{0.91~\mathrm{MeV}}\right),
    \label{eq:LDM_3}
\end{equation}
where $P$ is the thermal power of the nuclear reactor, which corresponds to 2.8~GW, and $E_{\gamma}$ is the $\gamma$-ray energy.
In the reactor core, the $\gamma$-rays can interact with electrons via a Compton-like process, resulting in the production of DPs~($\gamma+e\rightarrow A'+e$), as given by the following equation:
\begin{equation}
    \frac{dN_{A'}}{dE_{A'}}=\int\frac{dN_{\gamma}}{dE_{\gamma}}\frac{d\sigma_{\gamma e^-\rightarrow A'e^-}\left(E_{\gamma}\right)}{\sigma_{tot}dE_{A'}}dE_{\gamma},
    \label{eq:LDM_4}
\end{equation}
where $\sigma_{tot}$ represents the total interaction cross section between photons and matter in the reactor, and it is used for normalization.
The  $\gamma$-ray energy range utilized in the integration spans from 0.2 to 15 MeV.
In this production process, the DP flux is proportional to the square of the kinetic mixing parameter~($\varepsilon$) by $\sigma_{\gamma e^-\rightarrow A'e^-}$~\cite{deNiverville_2021}. 
Assuming $m_{A'}>2m_{\chi}$, the dark photon predominantly decays into two LDM particles~($A'\rightarrow\chi\chi$)~\cite{Filippi_2020}.
The strength of this decay mode is controlled by the dark fine structure constant $\alpha_D = g_{\chi}^2/4\pi \gg \alpha\varepsilon^2$, where $\alpha$ is the SM fine structure constant and $g_{\chi}$ is the dark coupling constant  between $A'$ and $\chi$. We used $g_{\chi} = 1$ for interpretation of LDM search, similarly to Ref.~\cite{Ge_2018}. In the DP interpretation, we use the case of  $\alpha_D = 0.1$ corresponding to $g_{\chi}$ = 1.1, close to 1. 

We consider signals induced by LDM scattering off electrons in the detector. 
The expected energy transfers from this process range from a few keV to a few tens of keV. The atomic structure of the orbiting electrons influences the measured energies. 
To account for the effect of energy transfer through the initial state bound electron to the final state outgoing electron, we use the atomic ionization factors obtained for sodium and iodine from Ref.~\cite{PhysRevD.108.083030}, which evaluates these factors from 0 to 10 keV.
Our region of interest (ROI) is defined between 1 and 10 keV. 

The expected scattered electron rate is derived as follows:
\begin{equation}
    \frac{dN_{e}}{dE_{e}}=\frac{N_{a}T}{4\pi R^2}\int\frac{2dN_{\chi}} {dE_{\chi}}\frac{d\sigma_{e}\left(E_{\chi}\right)}{dE_{e}}(IF_{\mathrm{Na}}+IF_{\mathrm{I}})dE_{\chi},
    \label{eq:LDM3_5}
\end{equation}
where $N_a$ represents the number of NaI molecules per kilogram of NaI crystal, $T$ is the exposure time normalized to days, $R$ is the distance between the detector and the reactor core, $E_{e}$ is the electron recoil energy, and $IF_{\mathrm{Na}}$ and $IF_{\mathrm{I}}$ are the integrated ionization factors for sodium and iodine, respectively. 
The differential cross-section of LDM scattering off an electron $\left( \frac{d\sigma_{e}\left(E_{\chi}\right)}{dE_{e}} \right)$ is calculated based on Ref.~\cite{Giudice_2018}.
Assume $m_{A'}=3m_{\chi}$ for the \textit{invisible decay} channel, we search for LDM in the mass range from 1 keV/c$^2$ to 1000 keV/c$^2$.

The NEON experiment began data collection in April 2022 following improvements in crystal encapsulation~\cite{choi2024upgrade}, achieving stable physics data. 
Data collected until June 2023 was utilized in this study. 
The reactor operated at full power from April 2022 to September 2022 and from February 2023 to June 2023, taking reactor-on data. 
The reactor was inactive from September 2022 to February 2023 for regular maintenance and fuel replacement, during which reactor-off data was collected. 
There was approximately one and a half months of downtime due to an unexpected power outage beginning on August 4, 2022.
Stable operations resumed on September 23, 2022, following recovery from the high voltage power supply failure. 

Two readout channels were obtained; however, the anode channel spectra were primarily used  for the ROI of 1--10 keV. 
The dynode channel readout was employed to understand background components, as detailed in Ref.~\cite{park2024}. 
Only single-hit events that did not coincide with LS or other crystals were used for this analysis, while multiple-hit events were rejected but used for background understanding.

PMT-induced noise events predominantly occur at low energies, below a few keV~\cite{Adhikari:2021rdm,Adhikari_2021}. A boosted decision tree~(BDT)~\cite{Coadou_2022} was adopted to mitigate this noise.
The BDT employs a decision tree algorithm with multivariate inputs to establish a robust discriminator that separates scintillation events from noise events. Boosting enhances accuracy by increasing the weights of misclassified events.
To better characterize scintillation events, we developed a waveform simulation package instead of relying on scintillation-rich calibration data. The waveforms generated by this package described scintillation waveform of the NaI(Tl) crystals from single photoelectrons to reconstructed variables as discussed in Ref.~\cite{choi2024waveform}. 
We used the waveform simulation dataset as the scintillation signal samples for the BDT training, while the single-hit physics data were used for the PMT-induced noise samples. The BDT output from each crystal's training determined the event selection criteria for each crystal~\cite{choi2024waveform,Adhikari_2021}. Selection efficiency is approximately 65\% at 1\,keV and reaches 100\% above 3\,keV.

\begin{figure*}[!htb]
   \begin{tabular}{cc}
      \includegraphics[width=0.49\textwidth]{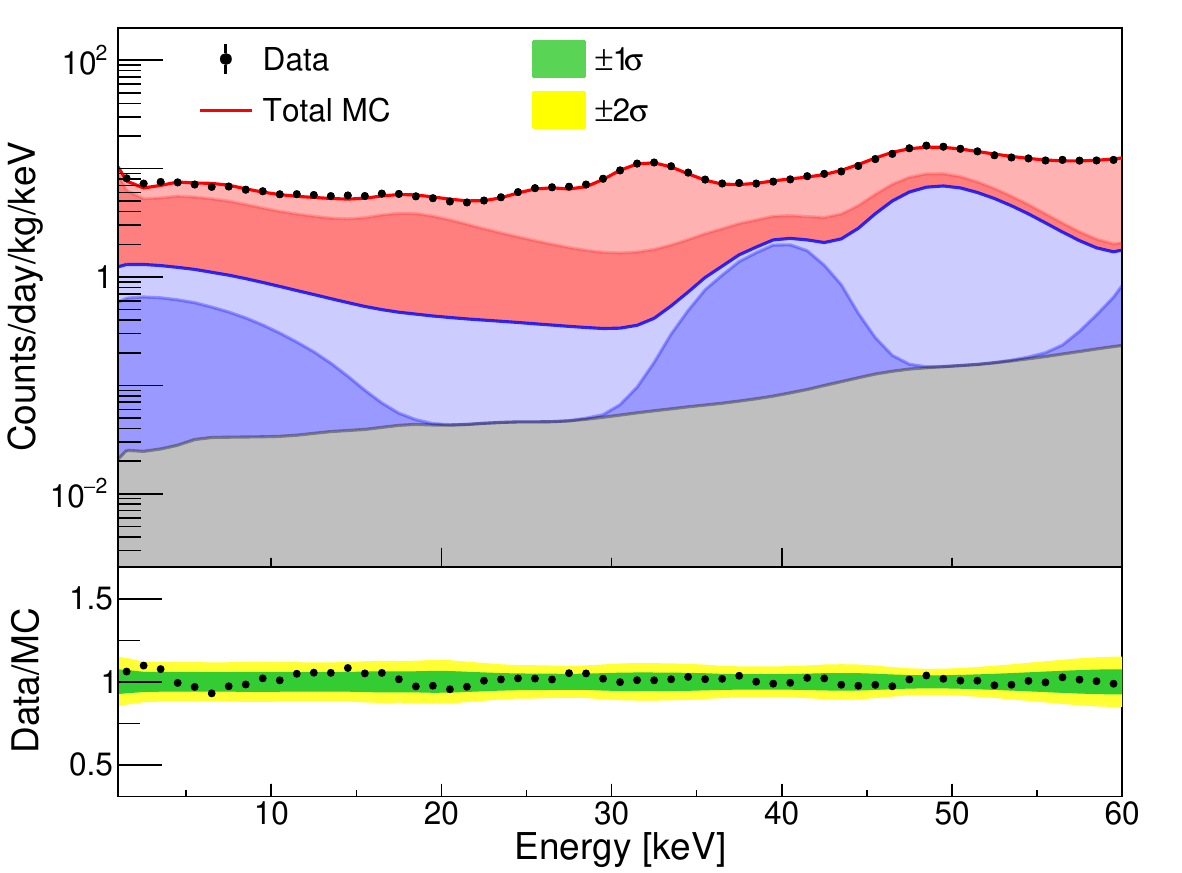} &
      \includegraphics[width=0.49\textwidth]{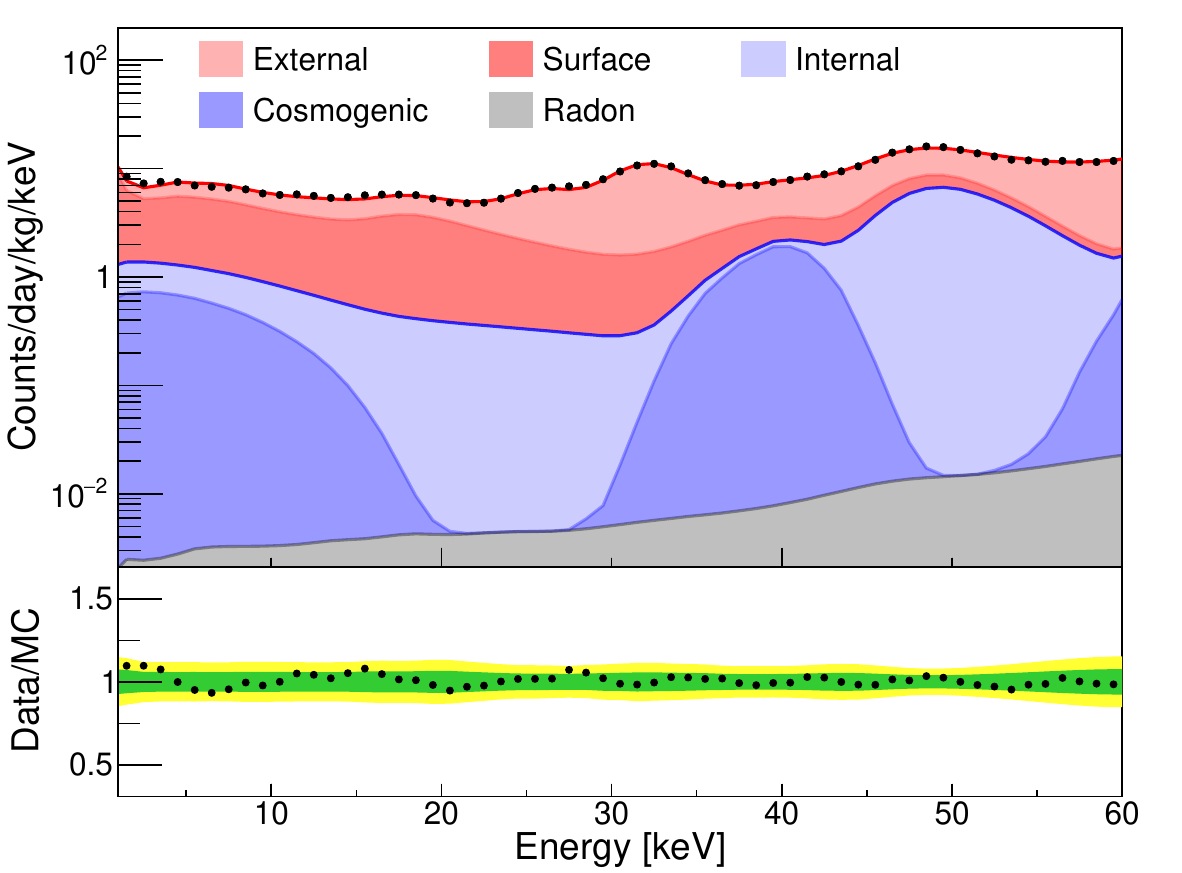}
    \end{tabular}
\caption{The energy spectra for reactor-on (left) and reactor-off (right) periods, combining data from five NEON detector modules, are shown. The spectra cover the energy range from 1\,keV to 60\,keV. Both spectra were modeled simultaneously, incorporating expected background components to accurately reflect the experimental conditions. Differences in background rates between reactor-on and reactor-off conditions are primarily attributable to seasonal fluctuations in radon levels. 
}
\label{fig:fig_1}
\end{figure*}

Low-energy event rate between 1\,keV and 3\,keV were monitored for each crystal after applying event selection criteria based on the BDT output.
Each one-hour dataset of each crystal that met the stable conditions within a 3$\sigma$ range of the overall rate was used for physics analysis.
Approximately 35\% of the data was excluded due to persistent PMT-induced noise events, with the entire dataset from detector-3 being excluded due to consistently high noise levels. 
The data exposure for this analysis amounts to 1,214\,kg$\cdot$days of reactor-on and 1,422\,kg$\cdot$days of reactor-off data.
Figure~\ref{fig:fig_1} shows the combined energy spectra from five crystal detectors during reactor-on (left) and reactor-off (right) periods.  

Background contributions in the NaI(Tl) detectors were analyzed using \Geant4-based Monte Carlo simulations. These simulations modeled measured data for both single-hit and multiple-hit events within the 3--3000\,keV energy ranges~\cite{park2024}. 
The simulation considered internal backgrounds, crystal surface or surrounding PTFE sheet contamination, cosmogenic activation, and external radiation. 
Additionally, we included the time-dependent backgrounds from $^{222}$Rn in the calibration holes, which have higher activities in Summer and lower in Winter, and initial dust contaminant in the liquid scintillator as detail elsewhere~\cite{park2024}. 
Because we removed the quartz window between the crystal and the PMT, external radiation from the PMT contributed more background to the NEON crystals compared to the COSINE-100 experiment~\cite{Adhikari:2021rdm}. This resulted in an  additional peak around 33\,keV from the K-shell dip~\cite{npr2024} and x-rays from Cs and Ba in the PMT photocathode~\cite{NAKAMURA2010276}.  As seen in Fig.~\ref{fig:fig_1}, the NEON data are well described by the expected backgrounds for both reactor-on  and reactor-off data. 

\begin{figure}[!htb]
\includegraphics[width=0.5\textwidth]{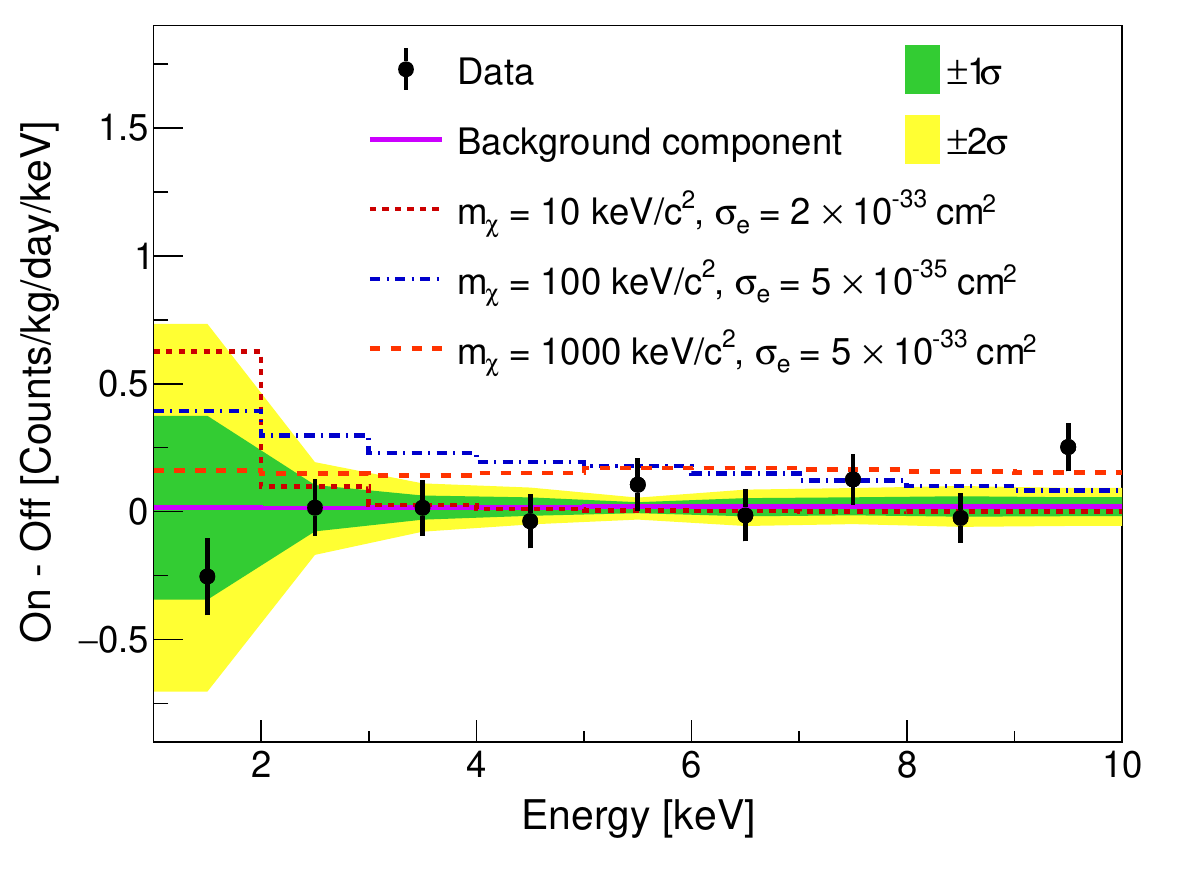} 
\caption{
The combined energy spectrum (black dots) from the five \NEON~detector modules is shown, along with the expected background (purple solid line) and $\pm1\sigma$ and $\pm2\sigma$ systematic uncertainty bands. For comparison, three benchmark signals were generated, assuming $m_{A'}=3m_{\chi}$ and $g_{\chi}=1$ for the \textit{invisible decay} channel, corresponding to LDM masses of 10\,keV/c$^2$ (dotted line), 100\,keV/c$^2$ (dashed-dotted line), and 1000\,keV/c$^2$ (dashed line).  
}
\label{fig:fig_2}
\end{figure}

Since most background contributions remained stable over the two-year data-taking period, we used reactor-on data subtracted by reactor-off data to isolate potential LDM signatures.
The time-dependent backgrounds, such as $^{222}$Rn in the calibration holes, dust contamination in the liquid scintillator, and cosmogenic activations due to cosmic muon radiation, were considered. 
Dominant time-dependent background is caused by seasonal variation of radon that was measured by the NEOS experiment~\cite{Ko_2017} and verified with the high energy data of the NEON experiment~\cite{park2024}. 
Figure~\ref{fig:fig_2} shows the reactor-on minus reactor-off spectrum derived from Fig.~\ref{fig:fig_1} with time-dependent background contributions. In the low-energy range, $^{222}$Rn contributions from the calibration holes are dominant but small enough to allow for the search for LDM signals. For comparison, three LDM interaction signal with masses of 10\,keV/c$^2$, 100\,keV/c$^2$, and 1000\,keV/c$^2$ are presented. 

Systematic uncertainties considered in this study include energy resolution, energy scale, selection efficiency, background component uncertainty, and exposure uncertainty. We accounted for these uncertainties by maximizing the difference between the reactor-on and reactor-off datasets, as shown in Fig.~\ref{fig:fig_2}. 
Among them, the selection efficiency driven by differences between reactor-on and reactor-off is the dominant systematic uncertainty, amounting to approximately 0.3 counts/kg/day/keV at 1 keV within the 1$\sigma$ level.

In the search for evidence of LDM-induced events, we performed a raster scan~\cite{lyons2014raster} using chi-square fits across 28 LDM mass values ranging from 1\,keV/c$^2$ to 1000\,keV/c$^2$. The measured data was fitted using crystal-specific background models and crystal-correlated signals for each crystal. The combined fit was achieved by summing the chi-square values from all five crystals. The chi-square depended on both signal strength and nuisance parameters, which control the background model to account for systematic uncertainties. Each nuisance parameter was constrained by its corresponding systematic uncertainty.

\begin{figure*}[!htb]
\includegraphics[width=0.98\textwidth]{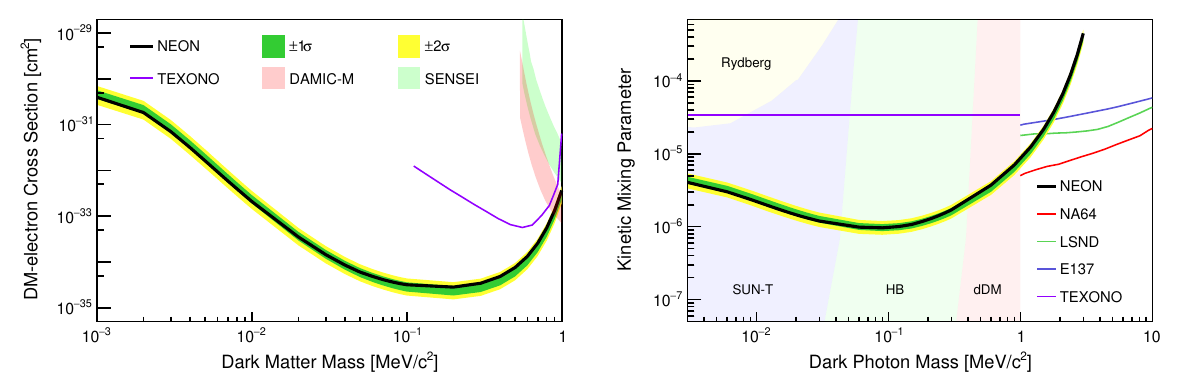} 
\caption{(Left) The 90\% confidence level exclusion limit on the LDM-electron scattering cross-section from NEON data assuming $m_{A'}=3m_{\chi}$ and $g_{\chi} = 1$, along with the $\pm 1\sigma$ and $\pm 2\sigma$ ranges of the expected sensitivities from 100,000 simulated experiments. This is compared with a phenomenological study of the TEXONO experiment~\cite{Ge_2018} and relic LDM search experiments of DAMIC-M~\cite{PhysRevLett.132.101006}, and SENSEI~\cite{Barak_2020}. 
(Right) The 90\% confidence level exclusion limit on the DP kinetic mixing parameter from NEON data via the LDM-electron scattering, along with of $\pm 1\sigma$ and $\pm 2\sigma$ expected sensitivities, is compared with the TEXONO result interpreted through the absorption process of dark photons~\cite{PhysRevLett.119.081801, Danilov_2019}. These DP parameter limits are compared with other limits derived from atomic experiment ($\mathrm{Rydberg}$)~\cite{Jaeckel_2010}, astrophysical observations (solar lifetime $\mathrm{SUN-T}$~\cite{Redondo_2013}, horizontal branch $\mathrm{HB}$~\cite{Hardy_2017}), diffuse X-ray background attributed to \dm~($\mathrm{dDM}$~\cite{Redondo_2009}), beam dump experiments ($\mathrm{E137}$~\cite{Batell_2014}, $\mathrm{LSND}$~\cite{PhysRevD.63.112001,PhysRevD.84.075020}, NA64~\cite{andreev2023search}) via \textit{ invisible decay} channel. Here $m_{A'}=3m_{\chi}$ and $\alpha_D = 0.1$, corresponding to $g_{\chi} = 1.1$, is used.}
\label{fig:fig_4}
\end{figure*}

We performed chi-square fits for the considered LDM masses. The minimum chi-square is $54.5$ at $m_\chi = 500~\mathrm{keV/c^{2}}$ and $\sigma_{e} = 4.84 \times 10^{-35}\,\mathrm{cm^{2}}$, while chi-square under the null hypothesis is $55.3$. The chi-square difference corresponds to a $p$-value of 67\%. 
The simulated datasets were generated with 100,000 iterations, accounting for systematic uncertainty, and were also utilized to estimate expected sensitivities. Based on these $p$-values, we determined that there is no significant LDM signal. A probability density function~(PDF) for each DM mass was defined using the combined chi-square values as follows:
\begin{equation}
    \mathrm{PDF} = C \exp \left(-\frac{\Delta\chi^2}{2}\right),
\end{equation}
where $C$ is a normalization factor that ensures the PDF integrates to 1 over the the physical region~($\sigma_e\geq0$). An upper limit was then estimated from the PDF at 90\% confidence level.

The resulting upper limits, shown in Fig.~\ref{fig:fig_4} (left), were compared with upper limits from a phenomenological study of the TEXONO experiment, assuming similar production and detection channels as in this analysis~\cite{Ge_2018}. Taking advantages of 1\,keV low energy analysis threshold compared to 3000\,keV of study for the TEXONO experiment, the NEON experiment significantly extended the mass range of the LDM search and greatly enhanced the exclusion limits. Our exclusion limits were also compared with relic LDM searches by DAMIC-M~\cite{PhysRevLett.132.101006} and SENSEI~\cite{Barak_2020}. Even though $N_{\mathrm{eff}}$ bounds limit the dark matter-electron scattering cross-section for dark matter masses below 10 MeV~\cite{B_hm_2013}, we did not include it to compare with terrestrial results. Despite being highly model-dependent, the NEON experiment's limits explored an extremely low mass region not previously reached by other laboratory experiments.

We also interpreted our limits in the $\varepsilon-m_{A'}$ parameter space assuming the \textit{invisible decay} of DP through the $\varepsilon$ extracted by cross-section of LDM scattering off an electron and the mass relationship between DP and LDM , as shown in Fig.~\ref{fig:fig_4} (right). This limit curve was compared with results from beam dump experiments~\cite{Batell_2014, andreev2023search} and the LSND experiment~\cite{PhysRevD.63.112001,PhysRevD.84.075020}, which consider the \textit{invisible decay} channel of DP, assuming $m_{A'}=3m_{\chi}$, consistent with this study. As demonstrated in the figure, reactor-based searches offer significant advantages for detecting low mass DP below 1000\,keV/c$^2$, a region inaccessible to accelerator-based experiments. A model-dependent comparison with TEXONO experimental data, considering the dark photon absorption  process in the detector~\cite{PhysRevLett.119.081801, Danilov_2019}, shows that our limits provide significantly better constraints for DP masses below 1000\,keV. The limits set by NEON data rule out parameter space already excluded by cosmological or astrophysical bounds for DP masses below 1000\,keV~\cite{Jaeckel_2010,Redondo_2009,Hardy_2017}. However, certain model-dependent environmental effects could potentially circumvent these limits. Importantly, this marks the first laboratory experiment to reach this parameter space, assuming the \textit{invisible decay} of DP.

In summary, we conducted a search for light dark matter (LDM) using 2,636\,kg$\cdot$days NaI(Tl) exposure data from the NEON experiment, located 23.7 meters from a 2.8 GW thermal power nuclear reactor. Our analysis established new exclusion limits for the dark matter-electron scattering cross-section ($\sigma_e$) for LDM masses ranging from 1\,keV/c$^2$ and 1000\,keV/c$^2$. This work significantly extends the LDM search range in laboratory experiments, achieving a 90\% confidence level upper limit of $\sigma_e$ = $3.17 \times 10^{-35}\,\mathrm{cm^{2}}$ for an LDM mass 100\,keV/c$^2$, marking the first experimental exploration into this parameter space. 
The NEON experiment continues to obtain reactor-on data, with more than twice the amount of data already collected compared to the current study. An updated study with a larger data sample and improved analysis techniques, focused on reducing the energy threshold, will allow us to explore a much larger region of
the currently unexplored LDM parameter space.


\acknowledgments
We thank Seokhoon Yun, Young-Min Lee and Luis~E.~Fran{\c c}a for insightful discussions. We thank the Korea Hydro and Nuclear Power (KHNP) company for the help and support provided by the staff members of the Safety and Engineering Support Team of Hanbit Nuclear Power Plant 3 and the IBS Research Solution Center (RSC) for providing high performance computing resources. This work is supported by the Institute for Basic Science (IBS) under Project Code IBS-R016-A1 and the National Research Foundation (NRF) grant funded by the Korean government (MSIT) (NRF-2021R1A2C1013761 and NRF-2021R1A2C3010989), Republic of Korea.


\bibliographystyle{PRTitle}
\bibliography{refer}

\end{document}